\documentclass[submission, Proceedings]{SciPost}

\hypersetup{
    colorlinks,
    linkcolor={red!50!black},
    citecolor={blue!50!black},
    urlcolor={blue!80!black}
}

\usepackage[bitstream-charter]{mathdesign}
\DeclareSymbolFont{usualmathcal}{OMS}{cmsy}{m}{n}
\DeclareSymbolFontAlphabet{\mathcal}{usualmathcal}

\newcommand{\w}{\omega}
\newcommand{\eq}[1]{eq.~\eqref{eq:#1}}

\begin{document}

\begin{center}{\Large \textbf{
Towards DIS at N4LO
}}\end{center}

\begin{center}
A. Basdew-Sharma\textsuperscript{1$\star$}
\end{center}

\begin{center}
{\bf 1} Nikhef, Amsterdam, The Netherlands
\\
* avanishb@nikhef.nl
\end{center}

\begin{center}
\today
\end{center}


\definecolor{palegray}{gray}{0.95}
\begin{center}
\colorbox{palegray}{
  \begin{tabular}{rr}
  \begin{minipage}{0.1\textwidth}
    \includegraphics[width=22mm]{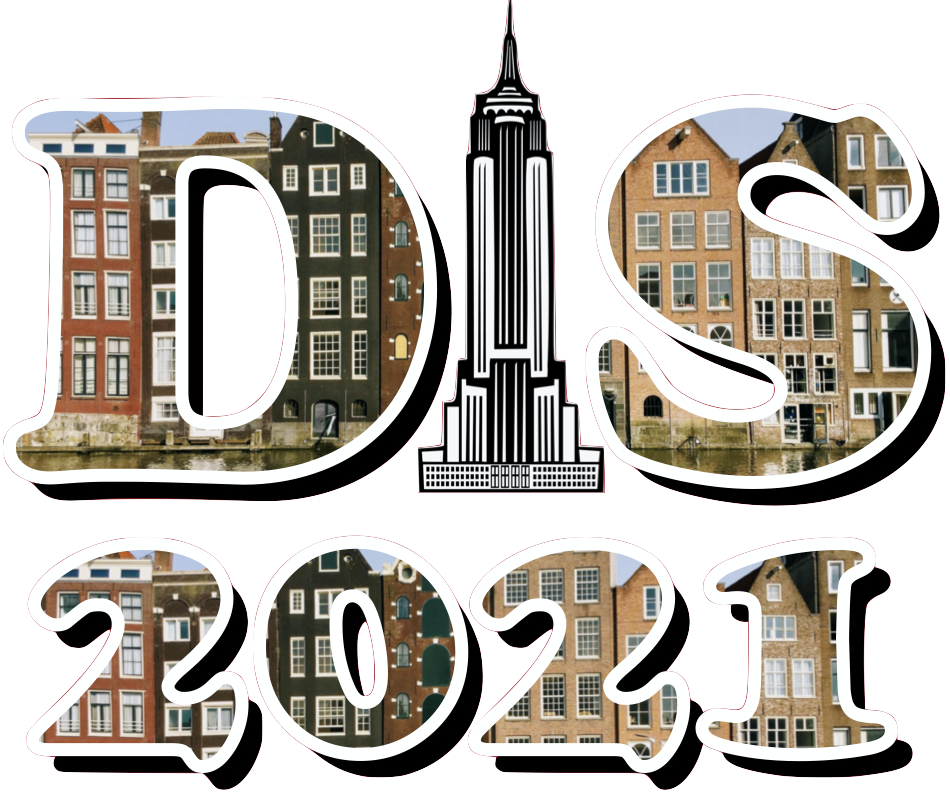}
  \end{minipage}
  &
  \begin{minipage}{0.75\textwidth}
    \begin{center}
    {\it Proceedings for the XXVIII International Workshop\\ on Deep-Inelastic Scattering and
Related Subjects,}\\
    {\it Stony Brook University, New York, USA, 12-16 April 2021} \\
    \doi{10.21468/SciPostPhysProc.?}\\
    \end{center}
  \end{minipage}
\end{tabular}
}
\end{center}

\section*{Abstract}
{\bf
We will report on an ongoing effort towards calculating the N4LO perturbative QCD corrections to the DIS total inclusive cross-section. We are developing a method based on differential equations and series expansion in the inverse Bjorken parameter. As a byproduct our calculation should also deliver analytic or at least precise numerical approximations for the four-loop splitting functions.
}

\vspace{10pt}
\noindent\rule{\textwidth}{1pt}
\tableofcontents\thispagestyle{fancy}
\noindent\rule{\textwidth}{1pt}
\vspace{10pt}

\section{Introduction}
\label{sec:intro}
With the forthcoming high luminosity upgrade at the Large Hadron Collider (LHC), the experimental uncertainty of the production of Higgs particles is expected to reduce from $\Delta_{\sigma_{H}}\sim 8\%$ (run 2) to $\Delta_{\sigma_{H}}\sim 3\%$ \cite{Amoroso:2020lgh}. To achieve the desired theoretical uncertainty for the predictions, one needs to perform calculations at N3LO precision in QCD perturbation theory. These calculations, in turn, formally require the four-loop splitting functions, which can be extracted from deep inelastic scattering (DIS) calculations at N4LO. However, due to the nonexistence of the four-loop splitting functions, predictions for a number of key observables have been calculated with the three-loop splitting functions \cite{Anastasiou:2015vya,Dreyer:2016oyx}, resulting in a higher theoretical uncertainty.
\par
Secondly, one can imagine that it will also be desirable to obtain more precise parton distribution functions (PDFs). Most of the data used to describe PDFs comes from DIS \cite{Accardi:2016ndt}. Therefore, to get a better idea of the PDFs, we need a better extraction of DIS data, for which N4LO DIS calculations can, at least, be useful. Moreover, the scale-evolution of the PDFs is governed by the splitting functions \cite{Accardi:2016ndt,Alekhin:2017kpj}, which is another incentive to extract splitting functions at N4LO, if one wants to reduce the theoretical uncertainty of the PDFs. 
\par
We are developing a new method to calculate the N4LO perturbative QCD corrections to the DIS total inclusive cross-section. As a byproduct we will be able to extract the four-loop splitting functions. The method is based on differential equations and series expansion in the inverse Bjorken parameter ($\w$). In particular, we will calculate the Mellin moments in a recursive manner, to compute coefficient functions from which we can build the structure functions and extract splitting functions. 
\par 
In the past the full $\w$ dependance of the coefficient functions was obtained from a finite number of Mellin moments. To mention some results: in 1999 Moch and Vermaseren calculated the DIS structure functions at two loops \cite{Moch:1999eb}. In 2004 Moch, Vermaseren and Vogt calculated the three-loop splitting functions in the non-singlet \cite{Moch:2004pa} and singlet case \cite{Vogt:2004mw}. After that it took Moch et al thirteen years to calculate the four-loop non-singlet splitting functions using an operator product expansion method \cite{Moch:2017uml}. In 2018 numerical approximations for the singlets were obtained using moments with $n<10$ \cite{Vogt:2018miu}. In the same year Herzog et al calculated the $n=2,3$ moments of the non-singlets at five-loops using the R* method \cite{Herzog:2018kwj}.
\par 
In our method we exploit the analytic behaviour of the moments at $\w\to 0$ to calculate expansions of the master integrals in $\w$, by using the differential equations obtained from the Integration-By-Parts (IBP) reduction tables (with \textit{Kira} \cite{Klappert:2020nbg} and FIRE \cite{Smirnov:2019qkx}, including \textit{LiteRed} \cite{Lee:2008tj,Lee:2012cn,Lee:2013mka}). We use FORM \cite{Vermaseren:2000nd,Kuipers:2012rf,Tentyukov:2007mu} to perform the symbolic manipulations and FORCER \cite{Ruijl:2017cxj} to obtain our boundary conditions of the differential equations. Our ultimate goal is to calculate the N4LO DIS QCD corrections and to obtain the four-loop splitting functions. However, in order to do so, we first need to validate our algorithm. To that end our short term goal is to first check known results at three-loop which have been checked before \cite{Ablinger:2010ty,Ablinger:2014vwa,Ablinger:2014nga,Ablinger:2017tan}, as well as unchecked results like the polarised splitting functions. 
\par
This report is structured in the following way: in section \ref{sec:setup} we will sketch the basic setup. In section \ref{sec:mellin} we will reconstruct the coefficient function from a finite number of Mellin moments. After that we will describe our recursive algorithm in section \ref{sec:rec} and we wil list some intermediate results in section \ref{sec:results}. We will conclude in section \ref{sec:conclusion}.

\section{Setup}
\label{sec:setup}
\begin{figure}[h]
\centering
\includegraphics[width=0.5\textwidth]{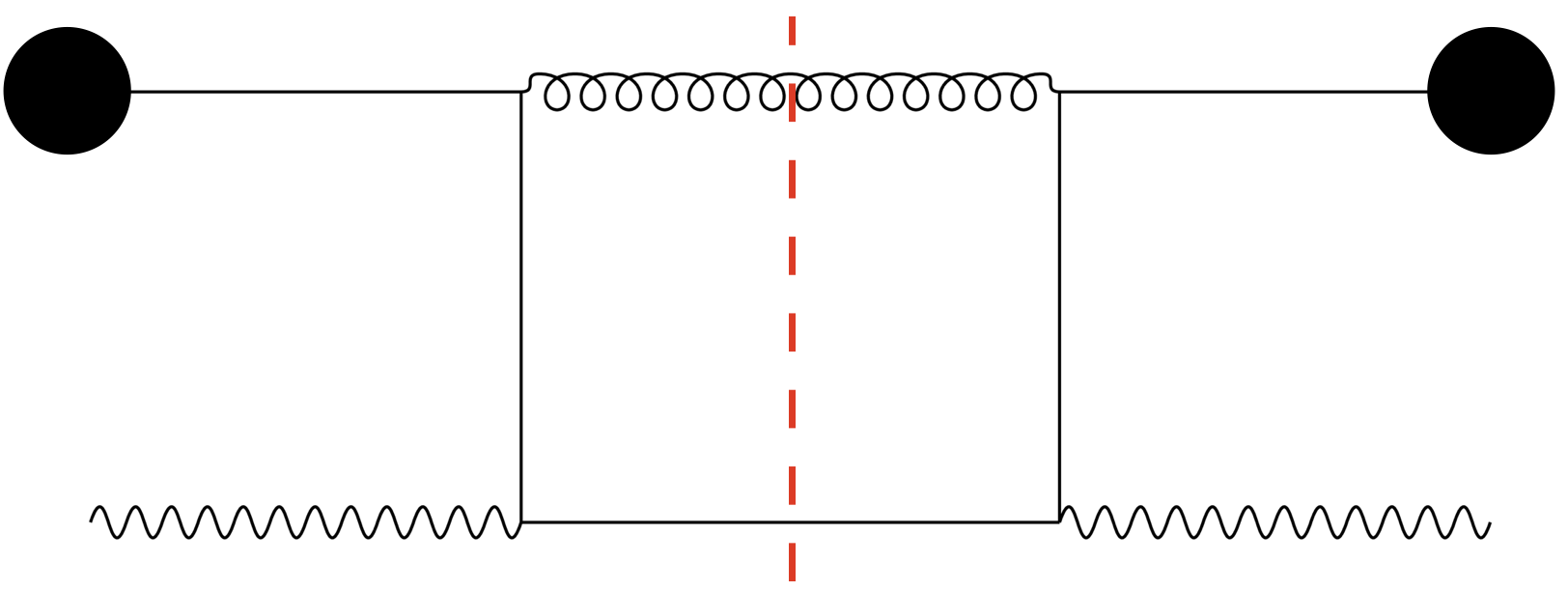}
\caption{Depicted we see the one loop DIS hadronic tensor, with $p$ and $q$ the momenta of the hadron and photon. The black dots are the hadrons and the red line is a cut.}
\label{fig:hadronictensor}
\end{figure}

\begin{figure}[h]
\centering
\includegraphics[width=0.5\textwidth]{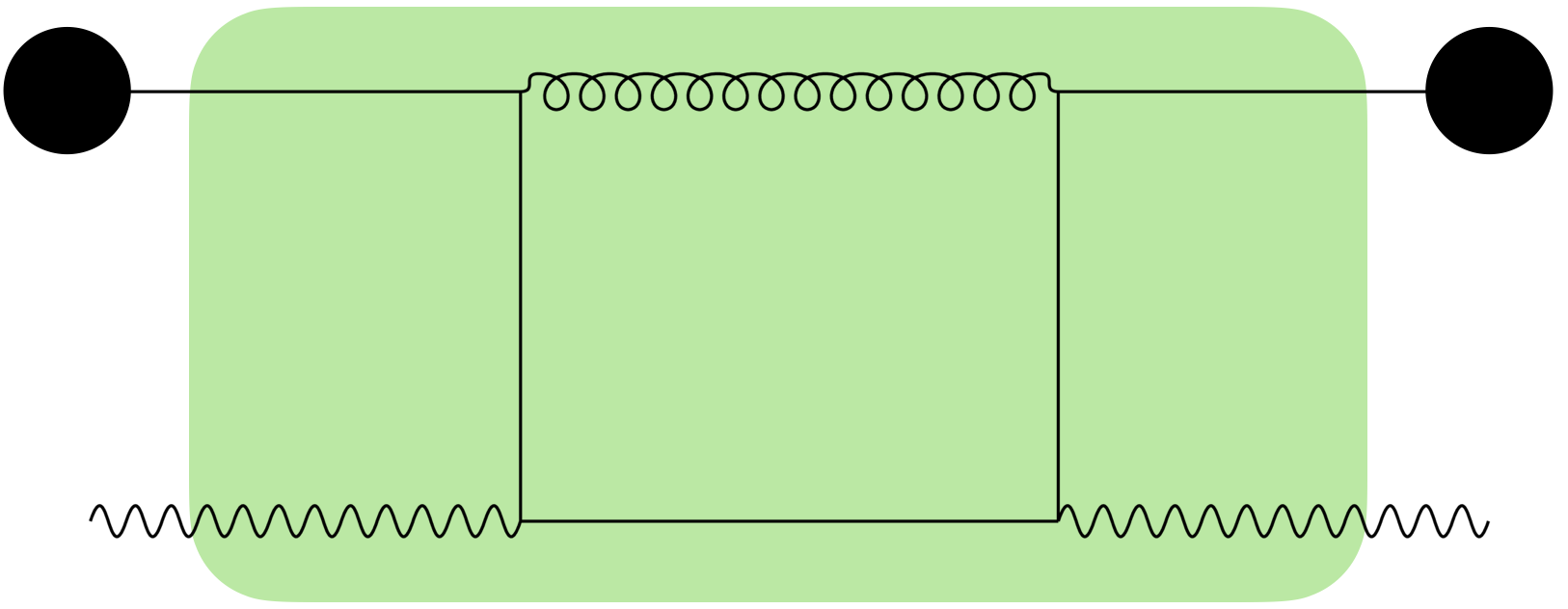}
\caption{To build the coefficient functions, we focus on the partonic part of the forward scattering diagram, highlighted in green.}
\label{fig:coeffun}
\end{figure}

This section is largely based on \cite{Sterman:1993hfp} and \cite{Bonocore:2016doa}. Our starting point will be the hadronic tensor $W_{\mu\nu}$ as sketched in figure \ref{fig:hadronictensor}.
By using the optical theorem, one can relate the hadronic tensor to the forward scattering amplitude $T_{\mu\nu}$ as 
\begin{equation}
W_{\mu\nu}(p,q)=2 \text{ Im } T_{\mu\nu}(p,q),
\label{eq:1}
\end{equation}
effectively removing the cut. In the following we will focus on the partonic level of the forward scattering amplitude to calculate the coefficient functions, as illustrated in figure \ref{fig:coeffun}. It is known that the structure functions can be constructed from the coefficient functions \cite{Vermaseren:2005qc}. 

\section{Method}
\label{sec:method}
Having sketched the setup, we can now proceed to our method. First we will show how to relate the Mellin moments to the forward scattering amplitude, as explained in 
\cite{Sterman:1993hfp} and \cite{Bonocore:2016doa}. After that we will sketch our recursive algorithm to obtain the Mellin moments. 

\subsection{Mellin moments}
\label{sec:mellin}
\begin{figure}[h]
\centering
\includegraphics[width=0.5\textwidth]{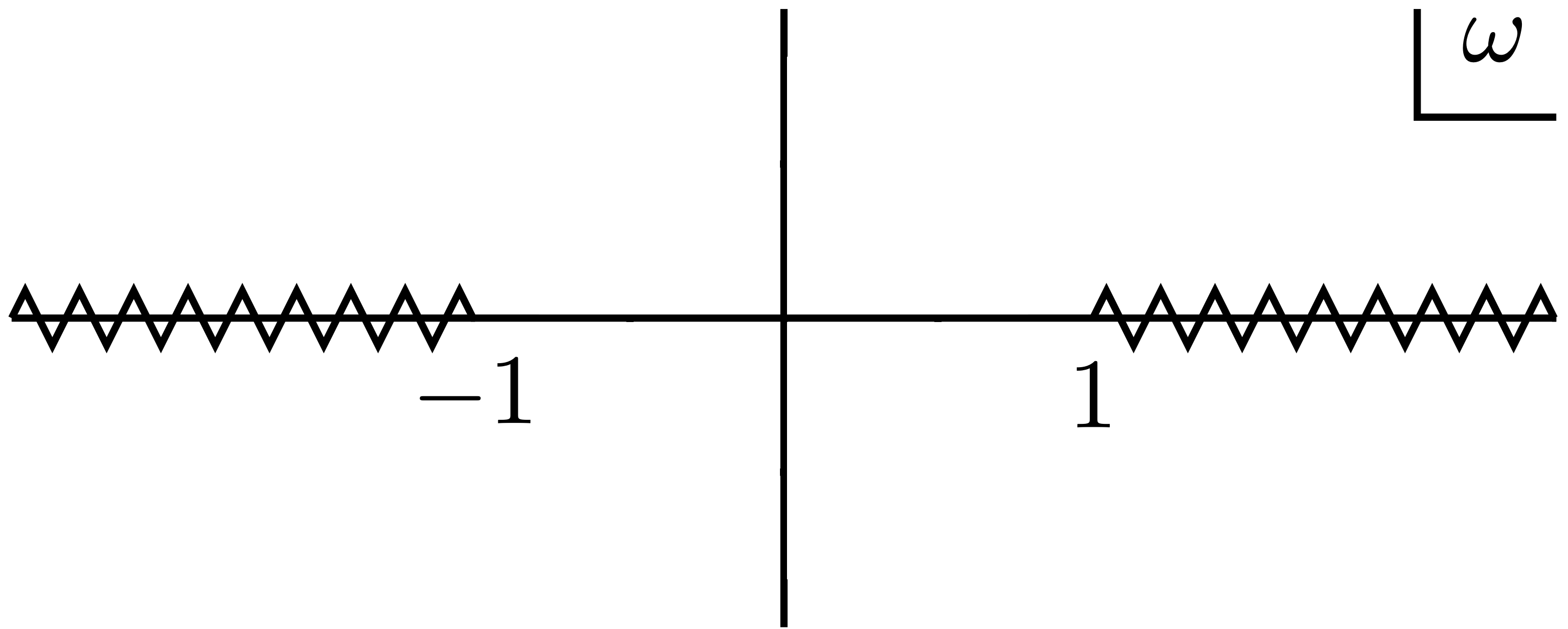}
\caption{The functions $T_{1}(\w,Q^{2})$ and $T_{2}(\w,Q^{2})$ have branch cuts for $\w\leq-1$ and $\w\geq1$.}
\label{fig:bc}
\end{figure}

\begin{figure}[h]
\centering
\includegraphics[width=0.8\textwidth]{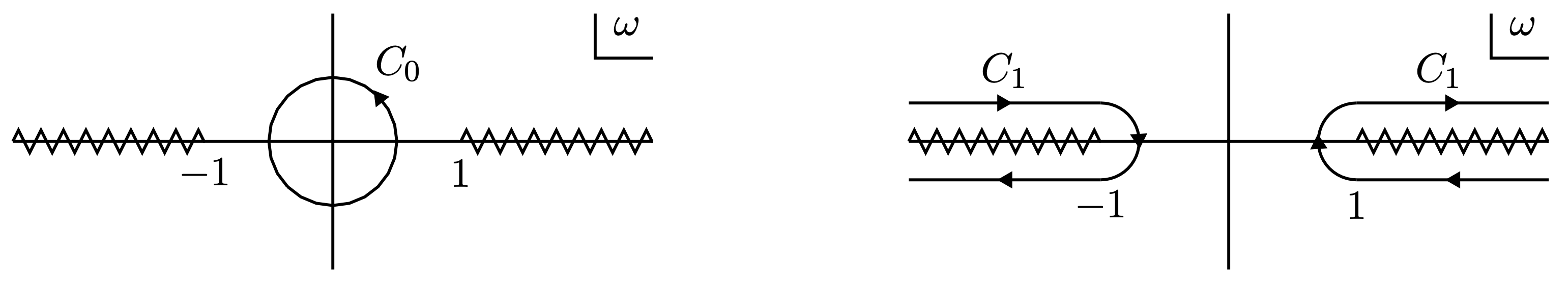}
\caption{The contour $C_{0}$ around the origin can be deformed into two contours $C_{1}$ around the branch cuts.}
\label{fig:contour}
\end{figure}

The forward scattering amplitude can be written as 
\begin{equation}
\label{eq:2}
T_{\mu\nu}(p,q)=-\left(\eta_{\mu\nu}-\frac{q_{\mu} q_{\nu}}{q^{2}}\right)T_{1}(\omega,Q^{2})+\left(p_{\mu}-\frac{q_{\mu}p\cdot q}{q^{2}}\right)\left(p_{\nu}-\frac{q_{\nu}p\cdot q}{q^{2}}\right)T_{2}(\omega,Q^{2}),
\end{equation}
where $\w\equiv x^{-1}$ is the inverse Bjorken parameter and $Q^{2} = - q^{2}$. The functions $T_{1}(\w,Q^{2})$ and $T_{2}(\w,Q^{2})$ have branch cuts for $\w\leq-1$ and $\w\geq1$ (figure \ref{fig:bc}).
Using Cauchy's theorem, one can relate the derivatives at $\w=0$ to a contour integral around it ($C_{0}$ in figure \ref{fig:contour}) as
\begin{equation}
\label{eq:3}
T^{(n)}_{i}(Q^{2})=\frac{1}{n!}\frac{d^{n}T_{i}(\omega,Q^{2})}{d\omega^{n}}\bigg|_{w=0}=\oint_{C_{0}}\frac{d\omega}{2\pi i}\frac{T_{i}(\omega,Q^{2})}{\omega^{n+1}}.
\end{equation}
However, the contour around the origin can be deformed into two contours ($C_{1}$ in figure \ref{fig:contour}) around the branch cuts.
Now we can use the fact that the functions $T_{1}$ and $T_{2}$ are symmetric in $\w$, such that we obtain 
\begin{equation}
\label{eq:4}
T_{i}^{(n)}(Q^{2})=\frac{(1+(-1)^{n})}{2\pi i}\int_{1}^{\infty}d\omega\frac{Disc_{\omega}T_{i}(\omega,Q^{2})}{\omega^{n+1}}.
\end{equation}
Notice that the discontinuity across the branch cuts is the imaginary part of the forward scattering amplitude. Therefore, by again using the optical theorem for even $n$, we can go back to $x$-space to obtain
\begin{equation}
\label{eq:5}
T_{i}^{(n)}(Q^{2})=\frac{1}{\pi}\int_{0}^{1}dx\ x^{n-1}W_{i}(x,Q^{2})=\frac{1}{\pi}\mathcal{M}_{n}[W_{i}(Q^2)].
\end{equation}
Thus we see that from a finite number of Mellin moments, one can obtain the full $\w$ dependance of the coefficient functions. This has been used extensively in the past to calculate splitting functions \cite{Moch:1999eb,Moch:2004pa,Vogt:2004mw}. In the next section we will exhaust the analytic behaviour at $\w\to0$ to calculate the solutions to the loop integrals.

\subsection{Recursive algorithm}
\label{sec:rec}
In this section we will explain our algorithm to recursively obtain solutions to the loop integrals involved in the calculations. We will first tackle the one-loop problem and then generalise to higher orders. 

\subsubsection{One-loop}
\begin{figure}[h]
\centering
\includegraphics[width=0.8\textwidth]{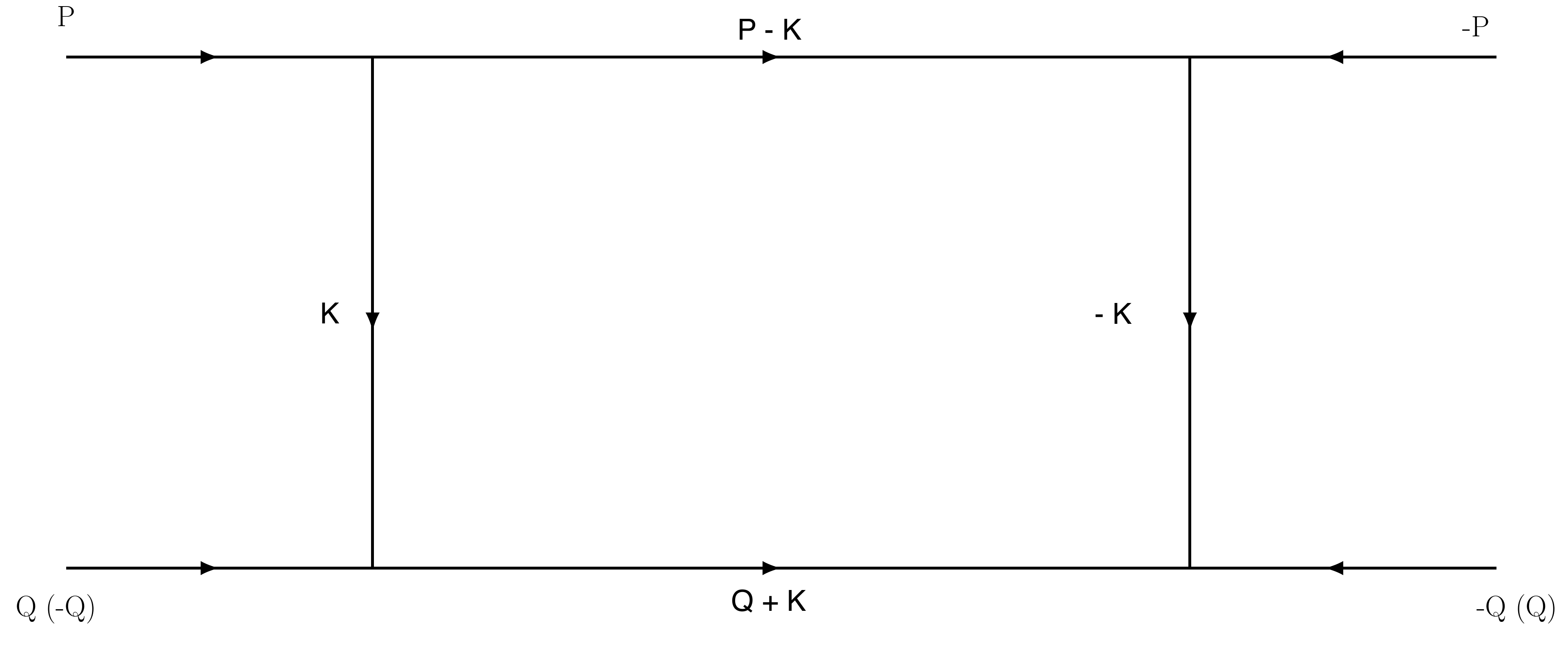}
\caption{The topology of topo1 is illustrated. Here we used the fact that one can exploit the symmetry $Q\leftrightarrow -Q$ to reduce the diagrams to a single topology.}
\label{fig:DISone}
\end{figure}

One can scalerize the loop integrals to sets of scalar integrals which belong different topologies. At one-loop order (see figure \ref{fig:DISone}) there is only one topology,
\begin{equation}
\text{Topo}(A1,A2,A3)=\int d^{D}K\left(\frac{1}{(K^{2})^{A1}}\right)\left(\frac{1}{((P-K)^{2})^{A2}}\right)\left(\frac{1}{((Q+K)^{2})^{A3}}\right).
\label{eq:6}
\end{equation}
Subsequently we used the program \textit{Kira} \cite{Klappert:2020nbg} to reduce all integrals in the topology to a set of master integrals, $M_{1}=\text{Topo}(1,0,1)$ and $M_{2}=\text{Topo}(0,1,1)$. Using the reduction tables, we can obtain a differential equation for $M_{2}$
\begin{equation}
\frac{\partial M_{2}}{\partial \omega} = -\frac{\epsilon}{1+\omega}M_{2}.
\label{eq:7}
\end{equation}
Notice that $M_{1}$ does not depend on $\w$. Our boundary condition for $M_{2}$ is $M_{2}(\omega\to 0)=M_{1}$. Next we use our series solution ansatz for the master integral
\begin{equation}
M_{2}(\omega,\epsilon)=\sum_{n=0}^{\infty}\omega^{n}a_{n}(\epsilon)
\label{eq:8}
\end{equation}
and notice that the left hand side of \eq{7} always is of a higher order in $n$ for fixed $\w$ than the right hand side. This allows us to derive a recursive relation for the coefficients 
\begin{equation}
a_{n} = -\frac{\left(\epsilon + n -1\right)a_{n-1}}{n}.
\label{eq:9}
\end{equation}
Thus, knowing the boundary conditions, one can build the full $\w$-dependent solutions to the master integrals.

\subsubsection{Multiple loops}
There arise three problems when one continues to higher order loops: 1. the boundary conditions should be calculated (which is hard in general), 2. one needs to deal with higher order poles in the differential matrix, and 3. the reductions will require a lot of computing power and runtime. 
\par
The boundary conditions are obtained by setting $\w\to0$. This is equivalent to taking $P\to0$. However, notice that the latter limit reduces all the diagrams to massless self energy diagrams. Luckily for us these have all been calculated (see \cite{Lee:2012cn,Lee:2013mka,Ueda:2016sxw,Ueda:2016yjm,Ruijl:2017eht,Baikov:2010hf,Lee:2011jt}) and stored by our group members using FORCER \cite{Ruijl:2017cxj}. Thus, all boundary conditions are known.
\par
For the second problem, let us consider a toy topology with three masters,
\begin{equation}
\frac{d}{d\omega}M_{i} = A_{ij}\ M_{j},
\label{eq:toy1}
\end{equation}
with
\begin{align}
\label{eq:toyA1}
A&=
\begin{pmatrix}
1 & 0 & 0\\
\frac{a}{\w^4} & 1 & 0\\
\frac{b}{\w^3} & 0 & \frac{c}{\w}
\end{pmatrix}.
\end{align}
Notice the higher order poles in the differential matrix $A$. We will transform these away order by order by using a scaling transformation $T$ 
\begin{align}
\begin{split}
\vec{M}'&=T\vec{M}\\
A'&=TAT^{-1} - T\partial_{\omega}T^{-1}.
\end{split}
\label{eq:scalings}
\end{align}
$T_{ij}$ is a diagonal matrix in which the $\text{i}^{th}$ entry equals the (negative) exponent in $\w$ of the highest order pole in the $\text{i}^{th}$ row in $A_{ij}$. Thus, to scale away the pole in the term $\frac{a}{\w^{4}}$ in equation \eq{toyA1}, $T$ is chosen as 
\begin{align}
\label{eq:T1}
T_{4}&=
\begin{pmatrix}
1 & 0 & 0\\
0 & \w^{4} & 0\\
0 & 0 & 1
\end{pmatrix},
\end{align}
such that 
\begin{align}
\label{eq:toyA2}
A'&=
\left(
\begin{array}{ccc}
 1 & 0 & 0 \\
 a & \frac{\w+4}{\w} & 0 \\
 \frac{b}{\w^3} & 0 & \frac{c}{\w} \\
\end{array}
\right).
\end{align}
Subsequently we have to scale away the $O(\w^-{3})$ term. We do this, by transforming with
\begin{align}
\label{eq:T2}
T_{3}&=
\begin{pmatrix}
1 & 0 & 0\\
0 & 1 & 0\\
0 & 0 & \w^{3}
\end{pmatrix},
\end{align}
resulting in 
\begin{align}
\label{eq:toyA3}
A''&=
\left(
\begin{array}{ccc}
 1 & 0 & 0 \\
 a & \frac{\w+4}{\w} & 0 \\
 b & 0 & \frac{c+3}{\w} \\
\end{array}
\right).
\end{align}
Therefore we are left with $O(\w^{-1})$ poles. As a remark we would like to mention that it is not possible to scale away the $O(\w^{-1})$ poles in the same manner. The final transformation matrix $T=T_{3}\cdot T_{4}$ becomes 
\begin{align}
\label{eq:Tfin}
T&=
\left(
\begin{array}{ccc}
 1 & 0 & 0 \\
 0 & \w^4 & 0 \\
 0 & 0 & \w^3 \\
\end{array}
\right).
\end{align}
As a consequence, according to \eq{scalings}, $M''_{2}$ and $M''_{3}$ start at order $O(\w^{4})$ and $O(\w^{3})$ respectively in their series solution. We conjecture that a diagonal transformation matrix is sufficient to map away deeper poles from the differential matrix. So far this has always worked, but a rigorous proof is lacking. 
\par
Lastly we need to deal with the $O(\w^{-1})$ terms left in the differential matrix. To that end it is convenient to pull out a factor of $\w^{-1}$ from the differential matrix, such that is it not singular anymore and obtain
\begin{equation}
\frac{d}{d\omega}M_{i} = \frac{A_{ij}}{\omega}M_{j}.
\label{eq:difmod}
\end{equation}
By expanding the masters and differential matrix as 
\begin{align}
\begin{split}
\label{eq:modrec2}
M_{i}=\sum_{k=0}^{\infty}c_{i}^{(k)}\w^{k},\\
A_{ij}=\sum_{k=0}^{\infty}A_{ij}^{(k)}\w^{k},
\end{split}
\end{align}
we can, by separating out the coefficient, build a new recursive relation for the coefficients $c_{l} $ of the master integrals as 
\begin{equation}
c_{l}^{(k)} = \left(B_{li}^{(k)}\right)^{-1}\sum_{0\leq k{'}\leq (k-1)}A_{ij}^{(k-k{'})}c_{j}^{(k{'})},
\label{eq:recmod}
\end{equation}
in which
\begin{equation}
B_{ij}^{(k)}\equiv \left(k\ \delta_{ij}-A_{ij}^{(0)}\right).
\label{eq:bmat}
\end{equation}
We would like to stress that this is a recurrence relation: for a fixed $l$ the order $k$ on the left hand side of \eq{recmod} is always larger than $k'$ on the right hand side. We have implemented this relation in a highly parallelized code in FORM \cite{Vermaseren:2000nd,Kuipers:2012rf,Tentyukov:2007mu}.

\section{Results}
\label{sec:results}
In this section we will list some intermediate results to display the performance of the algorithm.
\begin{center}
\begin{tabular}{ |c|c| } 
 \hline
 \#loops & \#topos \\ 
 \hline
 1 & 1 \\ 
 2 & 3 \\
 3 & 32 \\
 4 & 364 \\  
 \hline
\end{tabular}
\end{center}
The number of topologies at each loop ramps up quite fast from three loops and onward. 
\begin{center}
\begin{tabular}{ |c|c| } 
 \hline
 \#loops & time \\ 
 \hline
 1 & 3.7s (8 cores) \\ 
 2 & 90s (24 cores) \\
 3 & ~ week (64 cores) \\
 4 & ? \\  
 \hline
\end{tabular}
\end{center}
The reduction time of the topologies at each loop is small at one and two loop, however it increases to a week at three loops. We have not run the four-loop reduction yet.
\begin{center}
\begin{tabular}{ |c|c| } 
 \hline
 \#loops & time \\ 
 \hline
 1 & 2 \\ 
 2 & 31 \\
 3 & 2989 \\
 4 & ? \\  
 \hline
\end{tabular}
\end{center}
For the number of masters at each loop we see a similar pattern. The number of masters increases by two orders of magnitude at three loops. 
\par
Our toughest calculation for the coefficients so far is for topology 31 at three loops with 156 masters. It took ten hours to calculate 100 coefficients on a computer using 58 cores. It is uncertain whether we will be able to run the full four-loop reduction on our machine (64 cores, 1TB RAM memory). At least feasible should be $n_{f}^2$ pieces. We are in the process of discussing possible improvements of the reduction with one of the \textit{Kira} authors J. Usovitsch.

\section{Conclusion}
\label{sec:conclusion}
We have developed an algorithm to calculate the DIS cross-section at four loops. To introduce our new algorithm we first explained our setup to calculate the coefficient functions. We related the Mellin moments to the forward scattering amplitude. We showed how to obtain differential equations for the resulting master integrals, using IBP reduction tables. At one loop this was sufficient to derive a recursive relation. At higher orders in perturbation theory, we needed to cope with extra poles in $\w$ in the differential matrix. To this end we scaled away the poles, by transforming the master integrals and we modified the recurrence relation.
\par
To conclude, our algorithm has been checked to work at one- and two-loop orders. We are finishing up the calculation at three loops. We will start the four-loop reductions for the $n_{f}^2$ pieces soon.

\section*{Acknowledgements}
I would like to thank Andreas Vogt, Jos Vermaseren, Andreas Pelloni and Franz Herzog for their collaboration on this project.

\paragraph{Funding information}
This work is supported by the NWO Vidi grant 680-47-551.



\bibliographystyle{SciPost_bibstyle} 
\bibliography{DIS_proceedings_avanish_basdew-sharma_biblio}

\nolinenumbers

\end{document}